\documentclass[]{aa}  
\usepackage{orcidlink}

\usepackage{graphicx}
\usepackage{txfonts}
\begin{document}

\title{Extremely diverse coronal jets accompanying  an erupting filament captured by Solar Orbiter}
   \author{
          Song Tan
          \inst{1,2}~\orcidlink{0000-0003-0317-0534}
          \and
          Alexander Warmuth\inst{1}\thanks{Corresponding author; awarmuth@aip.de}
          \and
          Frédéric Schuller\inst{1}
          \and
          Yuandeng Shen\inst{3,4}
          \and 
          Jake A. J. Mitchell~\orcidlink{0000-0002-5493-1420}\inst{1}
          \and
          Fanpeng Shi\inst{5,1}
          }

   \institute{Leibniz-Institut für Astrophysik Potsdam (AIP), An der Sternwarte 16, 14482 Potsdam, Germany
   \and Institut für Physik und Astronomie, Universität Potsdam, Karl-Liebknecht-Straße 24/25, 14476 Potsdam, Germany
   \and State Key Laboratory of Solar Activity and Space Weather, School of Aerospace, Harbin Institute of Technology, Shenzhen 518055, China
   \and Shenzhen Key Laboratory of Numerical Prediction for Space Storm, Harbin Institute of Technology, Shenzhen 518055, China
   \and Key Laboratory of Dark Matter and Space Science, Purple Mountain Observatory, Chinese Academy of Sciences, Nanjing 210023, China}

   \date{Received/ accepted}

 
  \abstract
{Solar jets are collimated plasma ejections driven by magnetic reconnection, which play a critical role in the energy release and mass transport in the solar atmosphere. Using Solar Orbiter’s Extreme Ultraviolet Imager (EUI) with its unprecedented spatiotemporal resolution, we report the discovery of nine transient coronal jets associated with a filament eruption on September 30, 2024. These jets, with a median lifetime of only 22 seconds, have significantly shorter timescales than previously observed coronal jets. They exhibit diverse morphologies and properties, evolving through three distinct phases of the filament eruption: initiation, rise, and peak. The spatial and temporal distribution of the jets suggests they are driven by dynamic magnetic reconnection between the erupting filament and overlying magnetic fields. These jets represent a distinct class of phenomena different from traditional mini-filament-driven jets, being directly associated with large-scale filament eruption processes. This study reveals a previously unrecognised class of highly transient jets, highlighting the complexity of reconnection-driven processes during filament eruptions and underscoring the importance of high-resolution observations in uncovering fundamental plasma dynamics in the solar atmosphere.}


   \maketitle
%

\section{Introduction}

\begin{figure*}
    \centering
    \includegraphics[scale=0.65]{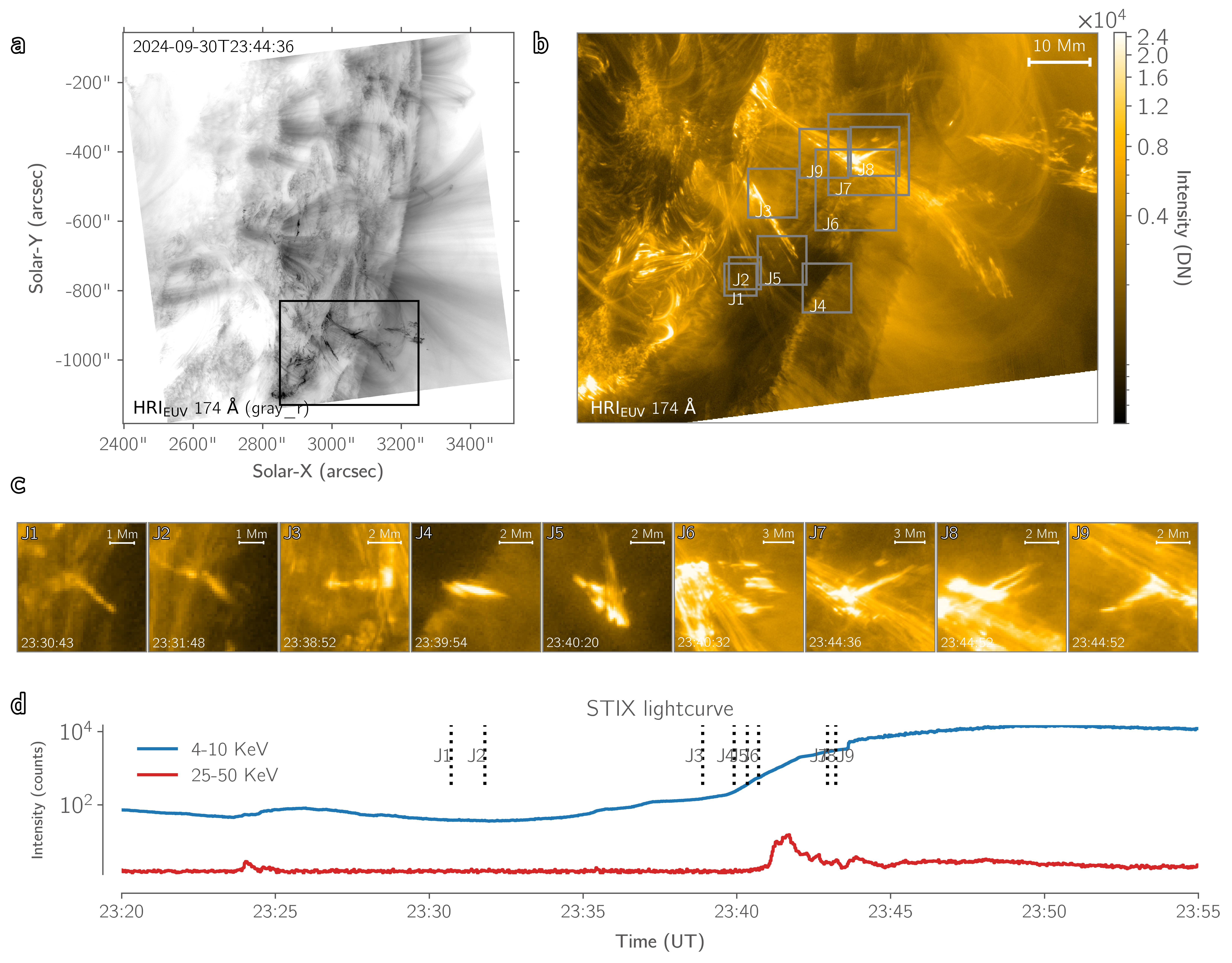}
    \caption{(a) $\mathrm{HRI_{EUV}}$ image (inverted colour scale) showing the erupting filament marked by a black rectangular box, with a length of approximately 80 Mm. (b) Magnified view of the erupting filament in $\mathrm{HRI_{EUV}}$, with gray rectangular boxes marking the locations of jets. (c) Most of the prominent frames of the nine jets, corresponding to the locations marked in (b). (d) STIX light curves in 4--10 keV and 25--50 keV bands (the background intensity was not subtracted), with the occurrence times of the nine jets indicated. It is worth clarifying that due to the intervention of the attenuator, the 4--10 keV light curve shows an abrupt decay from 23:44 UT. For this segment, we use the BKG detector data instead. An accompanying movie is provided online.}    
    \label{fig1}
\end{figure*}

The solar atmosphere is filled with dynamic activities across various scales, including jets, which are a significant manifestation of magnetic reconnection, playing a crucial role in our understanding solar activity, energy release, and mass transport processes \citep{2007Sci...318.1591S}. Solar jets, defined as collimated, beam-like plasma ejections along magnetic field lines, are ubiquitous in all regions of the solar atmosphere, including active regions, coronal holes, and quiet-Sun regions \citep{2016SSRv..201....1R,2021RSPSA.47700217S,2021PhDT........23J}. Based on morphological characteristics \citep{1994xspy.conf...29S}, solar jets can be classified into straight anemone jets and two-sided-loop jets \citep{1995Natur.375...42Y,2013ApJ...775..132J,2017ApJ...845...94T,2018ApJ...861..108Z,2019ApJ...883..104S,2019ApJ...887..220Y,2020MNRAS.498L.104W,2022MNRAS.516L..12T,2023MNRAS.520.3080T,2024ApJ...964....7Y}. \cite{2010ApJ...720..757M} further categorised straight anemone jets into standard jets and blowout jets. Standard jets exhibit typical inverted-Y structures, while blowout jets display more complex features, such as additional bright points inside the base arch and blowout eruption of the base arch (often containing a twisted mini-filament). 

The launch of Solar Orbiter has brought important breakthroughs to solar jet research \citep{2020A&A...642A...1M}. Its Extreme Ultraviolet Imager (EUI) includes the Full Sun Imager (FSI) and two High Resolution Imagers ($\mathrm{HRI_{EUV}}$ and $\mathrm{HRI_{Lya}}$), enabling observations of the solar atmosphere with an unprecedented spatial resolution \citep{2020A&A...642A...8R}.  Using EUI's high-resolution observations, \cite{2021A&A...656L...4B} discovered numerous small-scale EUV bright points termed "campfires" in the quiet solar atmosphere, opening up a unique observation window for small-scale activity  in the Solar Orbiter era. In particular, $\mathrm{HRI_{EUV}}$ achieves a spatial resolution of approximately 100 km/pixel when Solar Orbiter is at its closest approach to the Sun, coupled with a temporal resolution of a few seconds, providing ideal conditions for studying small-scale, rapidly evolving coronal jets. EUI's high-resolution observations have enabled a series of studies on small-scale dynamic features in the quiet Sun, including jets, blobs, and moving brightenings. These discoveries have greatly enriched our understanding of the physical properties of jets \citep{2021A&A...656L..13C,2022A&A...660A.143K,2023A&A...673A..82H,2023Sci...381..867C,2024A&A...686A.279S,2024A&A...692A.119C,2024A&A...692A.236N,2025ApJ...979..195D,2025arXiv250413009Z}. However, these jet observations have mainly focused on quiet regions and coronal holes \citep{2023A&A...675A.110B} and we are presently lacking $\mathrm{HRI_{EUV}}$ observations of coronal jets associated with  filament eruptions.

In this study, we leverage EUI/$\mathrm{HRI_{EUV}}$'s unique observations of a limb filament eruption on September 30, 2024, combined with on-disk observations from the Atmospheric Imaging Assembly \citep[AIA;][]{2012SoPh..275...17L} on board the Solar Dynamics Observatory (SDO), to analyze multiple transient jets produced during different phases of the filament eruption. These extremely short-lived jets (with a median lifetime of 22 seconds) are typically overlooked in traditional observations; however, they can be clearly resolved with $\mathrm{HRI_{EUV}}$'s high spatiotemporal resolution.


\section{Observations and results}
On September 30, 2024, Solar Orbiter was at a distance of 0.29 AU from the Sun, with a 101\degr\ angle relative to the Sun-Earth line, allowing the target filament to be well observed by both Solar Orbiter and SDO (Appendix \ref{A1}). $\mathrm{HRI_{EUV}}$ continuously observed the target filament for 60 minutes from 22:55 to 23:55 UT, which included the duration of the entire filament eruption, lasting from 23:20 to 23:55 UT. With unprecedented high spatial and temporal resolution (105 km/pixel and 2s, respectively), the dynamic features of this limb filament and accompanying jets during eruption were recorded in detail. We used X-ray light curves from the Spectrometer/Telescope for Imaging X-rays \citep[STIX;][]{2020A&A...642A..15K,2023A&A...673A.142X}  as supplementary observational data. Simultaneously, multi-wavelength AIA observations provided an on-disk view of the filament, while line-of-sight (LOS) magnetograms from the Helioseismic and Magnetic Imager (HMI) supplied magnetic field information of the filament and its surrounding environment. This paper primarily analyzes the limb eruption based on $\mathrm{HRI_{EUV}}$ observations, supplemented with SDO observational data for necessary contextual information. The data availability and analysis methods are detailed in Appendix \ref{A2} and \ref{A3}. 
We note that parallel investigations using similar datasets have recently emerged in the literature \citep{2025arXiv250312235C,2025ApJ...985L..12G,2025ApJ...988L..65B}. These concurrent studies, while sharing datasets, explore different aspects of solar eruptions, thereby providing complementary insights to our findings.

\subsection{Filament eruption and associated jets}
Figure \ref{fig1}(a) shows the filament eruption at its peak moment captured by $\mathrm{HRI_{EUV}}$, with the black rectangular region magnified in Fig. \ref{fig1}(b) to display the unwinding erupting filament and jets extending from its spine. The 35-minute (23:20 to 23:55 UT) HRI observation can be clearly divided into two phases: the eruption initiation marked by reconnection between emerging flux loops below the filament, followed by slow rotational rise of the filament producing an impulsive flare. The STIX light curves (Fig \ref{fig1}(d)) reflect this entire process, with the lower energies dominated by thermal emission (4--10 keV, yellow curve) and the higher energies dominated by nonthermal emission (25--50 keV, blue curve) showing two distinct enhancements. The first relatively weak enhancement around 23:23 UT corresponds to energy release from the reconnection between the emerging flux loop and the filament, while the second, stronger enhancement around 23:42 UT coincides with the main flare observed in $\mathrm{HRI_{EUV}}$. Here we focus on analyzing the dynamic transient jet activities appearing around the erupting filament, which exhibit rich morphological features. We refer to the attached movie to better understand the process.

\begin{figure}
    \centering
    \includegraphics[width=\linewidth]{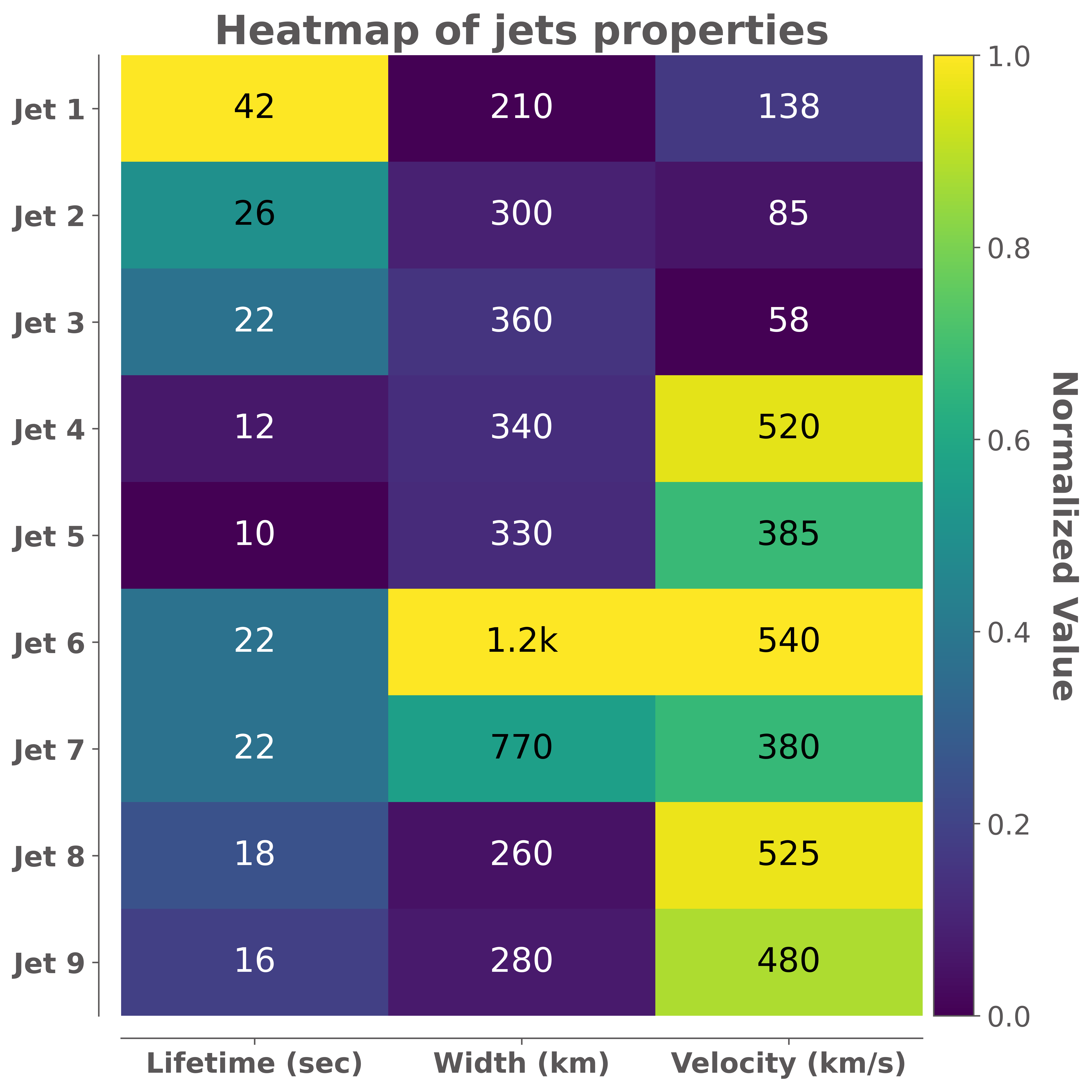}
    \caption{Heatmap of properties for the nine jets, displaying lifetime, width, and velocity parameters with absolute values represented numerically while relative magnitudes are indicated by colour intensity. The normalised value here refers to each property divided by the maximum value of that property among the nine jets, represented through the colour map, offering a way to intuitively reflect the variation trend of each property.
}
    \label{fig2}
\end{figure}

Through a careful frame-by-frame examination, we identified nine most prominent example jets (marked in Fig \ref{fig1}(b)) in the $\mathrm{HRI_{EUV}}$ images. These jets exhibited distinct collimated plasma outflows with clear spatial boundaries and temporal evolution separate from surrounding emission. Spatially, the nine jets appeared progressively with the rising height of the erupting filament, exhibiting a distribution clearly associated with the filament's evolution. Figure \ref{fig1}(c) shows a frame for each jet, while marking the corresponding occurrence times on the STIX light curve. The jets were produced following the initial X-ray light curve
 enhancement, with jets 1 and 2 occurring during the filament eruption initiation phase, and the rest concentrated in the peak eruption phase. Based on the spatial and temporal distribution of these jets, we inferred a strong correlation between them and the erupting filament, suggesting these jets are products of magnetic reconnection processes associated with the filament eruption.

\subsection{Jet properties and morphology}

Considering that jets in $\mathrm{HRI_{EUV}}$ images are very transient and dynamic, placing slices at jet locations to obtain time-space plots is not feasible. Therefore, we analyzed each jet frame-by-frame to obtain its lifetime, width, and velocity (Appendix \ref{A3} for specific analysis methods). Figure \ref{fig2} presents a heatmap that displays the property differences among the nine jets, with the normalised intensity of each property indicated using the colour scale.

Despite the limited sample size (nine jets with a total of 27 measurement points), several significant patterns emerge. As the filament eruption progresses, we observe a general trend toward shorter jet lifetimes and higher velocities. The width parameter shows more complex variation without a clear progressive pattern throughout the eruption sequence. Notably, the most distinct separation in parameter space appears in the velocity measurements, particularly between jets 3 and 4, which coincides with the transition from the initiation phase to the rising and peak eruption phase. Overall, these nine jets exhibit considerable variation in lifetime (median 22 s) and velocity (median 385 km/s), while their widths (median 330 km) demonstrate fewer systematic variations across the eruption process. However, it should be noted that because the jet show obvious bifurcation in the peak phase (noting that we only calculated the width of part of the bifurcation), the actual jet width properties also show an increasing trend.

Based on their temporal occurrence during different phases of the filament eruption and their physical characteristics, we classify these jets into three groups: group 1 (jets 1--3) corresponding to the filament eruption initiation phase; group 2 (jets 4 and 5) associated with the filament rising phase; and group 3 (jets 6--9) coinciding with the peak eruption phase. Jets within each group show high correlation in terms of properties and morphology, providing a foundation for selecting one representative jet from each eruption phase for detailed morphological analysis.

\begin{figure}
    \centering
    \includegraphics[width=\linewidth]{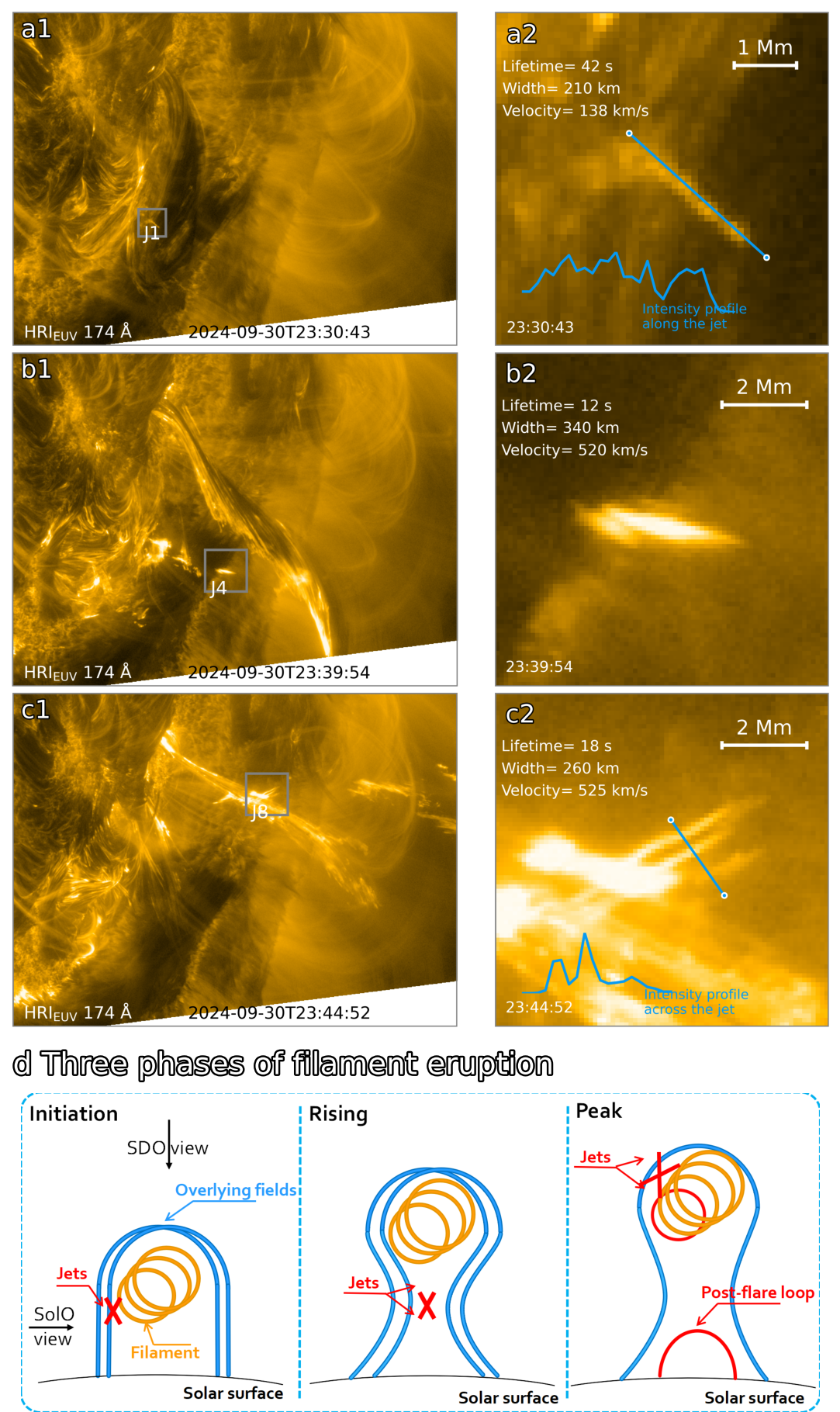}
    \caption{ Group 1 jet representative: Jet 1 exhibiting standard jet morphology, with its key properties labeled (a1 and a2). The blue curve shows intensity distribution along the jet. Group 2 jet representative: Jet 4 displaying standard jet morphology with its key properties labeled (b1 and b2). Group 3 jet representative: Jet 8 showing blowout jet characteristics with its key properties labeled (c1
and c2). The blue curve shows intensity distribution across the jet. Cartoon illustration explaining the jets produced during different phases of filament eruption (initiation, rising, peak), with annotations showing the viewing angles from Solar Orbiter and SDO (d)
.}
    \label{fig3}
\end{figure}

The left panels in Fig.~\ref{fig3} show $\mathrm{HRI_{EUV}}$ images containing the entire erupting filament, while the right panels present jets with corresponding parameters. Figures~\ref{fig3}(a1--c1) correspond to three stages of filament eruption (initiation, rising, peak), reflected in the changing height of the filament. Based on their morphology, \cite{1994xspy.conf...29S} classified coronal jets into straight anemone jets and two-sided-loop jets, explained as the result of emerging flux reconnecting with open and closed fields. \cite{2010ApJ...720..757M} further divided straight anemone jets into standard jets and blowout jets, with the significant feature of blowout jets being their association with erupting mini-filaments. Among our nine jets, those from the initiation phase (group 1) and rising phase (group 2) morphologically resemble standard jets, while jets from the peak eruption phase (group 3) match the blowout jet characteristics. The rising phase jets (group 2) exhibit significantly shorter lifetimes and stronger emission compared to initiation phase jets (group 1). Regarding locations, both initiation and peak phase jets (groups 1 and 3) are rooted on the erupting filament despite their property differences, while rising phase jets (group 2) are located below the rising filament, at some distance from the filament spine (compare Fig. \ref{fig3}(b1) with (a1), (c1)).

Case and statistical studies have found that mini-filament eruptions appear to be a common trigger mechanism for coronal jets despite morphological differences \citep{2011ApJ...735L..43S,2012ApJ...745..164S,2017ApJ...851...67S,2016ApJ...832L...7P,2017ApJ...844..131P,2018ApJ...853..189P,2012RAA....12..573C,2013RAA....13..253H,2013A&A...559A...1S,2015Natur.523..437S,2020ApJ...889..187S,2024ApJ...968..110D,2024ApJ...969...48P}. Similarly to the continuum of jet morphologies described by \cite{2015Natur.523..437S}, our observations also suggest that jets (group 1, narrow spire with a relatively dim base) from the initiation phase  exhibit standard-jet-like features when the reconnection is limited, while jets (group 3, wider spire and bright base) from the peak phase display blowout-jet-like features as the reconnection becomes more vigorous and extensive. This continuum of jet morphologies appears to depend on the evolutionary phase of the filament eruption. Additionally, we observed substructures in both radial (jet 1, intensity profile in Fig. \ref{fig3}(a2)) and transverse directions (jet 8, intensity profile in Fig. \ref{fig3}(c2)). Specifically, the three-pronged structure observed in jet 8 is hypothesised to represent the fine-scale structure of the erupting filament retained after its reconnection with the overlying fields (e.g., filament threads). The discussion below offers an   in-depth analysis of the generation mechanisms and substructures of these jets associated with different phases of filament eruption.

\section{Discussions}
\subsection{Unique spatiotemporal scales of the observed jets}
In a recent large-sample statistical study \citep{2024A&A...688A.127M}, 883 AIA EUV jets from 2011 to 2016 showed a median lifetime of 12.8 minutes, consistent with many previous studies \citep{2016A&A...589A..79M,2016ApJ...822L..23P,2020ApJ...889..183M,2021Ge&Ae..61.1083K}. However, among the nine observed jets, even the longest-lived (jet 1 from the initiation phase) did not exceed 1 minute. The median lifetime of our reported jets (22 seconds) is approximately 35 times shorter than those typically reported. Even disregarding spatial resolution limitations, jets with such short lifetimes would be unidentifiable with instruments having lower temporal resolution. With $\mathrm{HRI_{EUV}}$'s high temporal resolution of 2 seconds, jet 5 from the rising phase was captured in only 5 frames (Appendix \ref{A3}), indicating the significantly transient nature of our reported jets. Moreover, $\mathrm{HRI_{EUV}}$'s unique perspective and unparalleled spatial resolution were key factors enabling us to capture these transient features associated with the erupting filament. 
\cite{2023ApJ...943..156K} and \cite{2020ApJ...899...19C} both reported transient jet 
phenomena around filaments with properties remarkably similar to our observed jets. \cite{2023ApJ...943..156K} studied plasma blobs and jets around a failed filament eruption (width: 2--3 arcseconds, recurrence: 70 seconds), while \cite{2020ApJ...899...19C} analyzed tornado mini-jets from suspended prominences (lifetime: 10--50 seconds, width: 0.2--1.0 Mm, velocity: 100--350 km/s). These studies suggest that reconnection between the filament and overlying magnetic field may be the potential mechanism driving these jets. 
Recent numerical simulations combined with observational studies have demonstrated that high-resolution EUV imaging is essential for detecting small-scale coronal phenomena previously predicted by theory \citep{2023ApJ...943...24P,2024A&A...687A.171F}. Consistent with these predictions, \cite{2025arXiv250603092N}  reported the first detection of ultra-thin coronal jets (253--706 km wide by $\mathrm{HRI_{EUV}}$) and plasmoid-mediated reconnection signatures at spatial scales previously unresolvable by EUV imaging instruments, revealing new insights into fine-scale coronal dynamics. \cite{2024A&A...691A.198J} also used data from ground-based telescopes to demonstrate the fundamental role of high-resolution observations in determining the mechanism of jet generation.

Given the ubiquity of interactions between erupting filaments and overlying magnetic fields, we propose that these transient jet phenomena observed by $\mathrm{HRI_{EUV}}$ are likely widespread in the solar atmosphere but remain undetected as continuous moving jet
 structures due to the spatiotemporal resolution limitations.

\subsection{Jet generation mechanisms}
Mini-filament eruptions have been widely confirmed as a powerful mechanism driving coronal jets through observations and numerical simulations \citep{2015Natur.523..437S,2017Natur.544..452W}. However, the jets we observed differ significantly from these typical jets. While traditional jets originate from the photospheric and chromospheric base, our jets closely accompany the rising erupting filament and are spatially located in the corona. This characteristic resembles nanojets observed near coronal loops, which result from magnetic reconnection in a braided field \citep{2021NatAs...5...54A,2022ApJ...934..190S}.

SDO's on-disk view of the filament (Appendix \ref{A1}) revealed that one end of the filament was rooted near a sunspot region, indicating stronger overlying fields above this footpoint. We propose that reconnection between the erupting filament and these overlying fields directly triggered our observed jets (see the cartoon illustration in Fig. \ref{fig3}(d)). This interaction evolved through the eruption phases: beginning with relatively limited reconnection during the initiation phase (corresponding to group 1, jets 1--3), developing through the rising phase (group 2, jets 4 and 5), and reaching fully developed reconnection during the peak eruption phase (corresponding to group 3, jets 6--9). The progressive increase in jet velocities across these phases supports this scenario of intensifying reconnection throughout the eruption process. Typical coronal jets propagate along open field lines following reconnection between mini-filaments and open fields, with twist transfer from the filament to these open fields. Such open field configurations provide favorable conditions for jets to propagate into interplanetary space, along with a stable magnetic environment supports their sustained existence. In contrast, our observed jets accompanying the erupting filament were highly dynamic and transient, consistent with complex reconnection processes between the erupting filament and overlying magnetic fields. Most jets moved at significant angles to the filament spine and extended outside the filament boundary, which rules out the possibility that reconnection inside the filament triggered these jets. The rapid dissipation of these jets likely reflects the efficient conversion of magnetic energy into kinetic and thermal energy through the reconnection process.

Unlike jets from the initiation and peak phases (groups 1 and 3), which were directly associated with the filament, jets from the rising phase (group 2) were produced below the erupting filament (with obvious spatial distance from the filament spine, Fig. \ref{fig3}(b1)). These rising phase jets exhibited distinct bidirectional characteristics; for example, jet 4 moved upward with the erupting filament while jet 5 showed downward motion. By comparing the co-spatial features of post-flare-loops top and jet location (see the results of the 3D reconstruction in Appendix \ref{A2}), we suggest that reconnection of overlying magnetic fields, which were stretched during the filament eruption, drove the formation of these jets below the filament. This reconnection process naturally explains the observed bidirectional motion of the jets. 

\section{Conclusions}
Utilising the excellent spatiotemporal resolution of Solar Orbiter's $\mathrm{HRI_{EUV}}$  on September 30, 2024, we discovered a previously unreported class of transient coronal jets with a median lifetime of only 22 seconds, which is significantly shorter than traditionally observed coronal jets. Our main conclusions are reported below.    \begin{enumerate}
    \item Based on their association with filament eruption phases, we classified these jets into three groups (initiation, rising, and peak phases), revealing a systematic evolution toward higher velocities and shorter lifetimes as the eruption progresses. Their spatial distribution varies by phase: the initiation and peak phase jets are rooted directly in the filament spine, whereas rising phase jets appear below the erupting filament.

    \item We propose that dynamic reconnection between the erupting large-scale filament and overlying magnetic fields produces these transient jets, representing a mechanism distinct from traditional mini-filament-driven jets. This progression from relatively limited interactions during initiation to  fully developed reconnection at the peak phase explains the observed trends in jet properties.
   \end{enumerate}

Filaments, as widely present magnetised plasma structures in the solar atmosphere, exhibit rich dynamic characteristics \citep{2011LRSP....8....1C,2012LRSP....9....3W,2012ApJ...745L..18A,2015LRSP...12....3W,2014ApJ...786..151S,2014ApJ...795..130S,2023ApJ...955...87F,2024ApJ...968...85Z,2025ApJ...980...18F}. The connection between jets and large-scale filament eruptions has been explored
in previous studies, both theoretically \citep{2021ApJ...912...75L} and observationally \citep{2017SoPh..292...81C,2023A&A...672A..15J}. Previous research has also shown that mini-filament eruptions can drive coronal jets. Our findings establish a new perspective on the relationship between filaments and solar jets, where jets are also widely present in association with medium-to-large-scale erupting filaments, with their properties evolving systematically through the eruption phases. These transient jets reveal complex interaction patterns between erupting filaments and surrounding magnetic fields that had not been previously recognised in lower spatiotemporal resolution observations.

These discoveries provide new insights into magnetic reconnection dynamics during filament eruptions. The evolutionary pattern of jets we observe across the various phases of the filament eruption suggests that reconnection processes intensify throughout the eruption, with progressively stronger magnetic interactions manifesting as shorter-lived, faster jets as the eruption reaches its peak. With Solar Orbiter's continued operation, we expect to discover more similar phenomena, further revealing the fundamental physical processes of magnetic energy release and plasma acceleration in the solar atmosphere.

\begin{acknowledgements}
We thank the anonymous reviewer for constructive comments, which greatly improved the scientific quality and readability of our paper. S.T. thanks Hannah Collier of FHNW for her constructive suggestions and efforts in organising the observation plan.
Solar Orbiter is a space mission of international collaboration between ESA and NASA, operated by ESA. The EUI instrument was built by CSL, IAS, MPS, MSSL/UCL, PMOD/WRC, ROB, LCF/IO with funding from the Belgian Federal Science Policy Office (BELPSO); the Centre National d'Etudes Spatiales (CNES); the UK Space Agency (UKSA); the Bundesministerium f\"ur Wirtschaft und Energie (BMWi) through the Deutsches Zentrum f\"ur Luft- und Raumfahrt (DLR); and the Swiss Space Office (SSO). The STIX instrument is an international collaboration between Switzerland, Poland, France, Czech Republic, Germany, Austria, Ireland, and Italy. The AIP team was supported by the German Space Agency (DLR), grant numbers \mbox{50 OT 1904} and \mbox{50 OT 2304}. A.W. and J.M. also acknowledge funding by the European Union’s Horizon Europe research and innovation program under grant agreement No. 101134999 (SOLER). Y.S. was supported by the  Shenzhen Key Laboratory Launching Project (No. ZDSYS20210702140800001) and the Specialized Research Fund for State Key Laboratory of Solar Activity and Space Weather.
\end{acknowledgements}

\bibliographystyle{aa} 
\bibliography{ref} 

\begin{appendix}
\section{SDO on-disk observation}
\label{A1}
\begin{figure*}
    \centering
    \includegraphics[width=\linewidth]{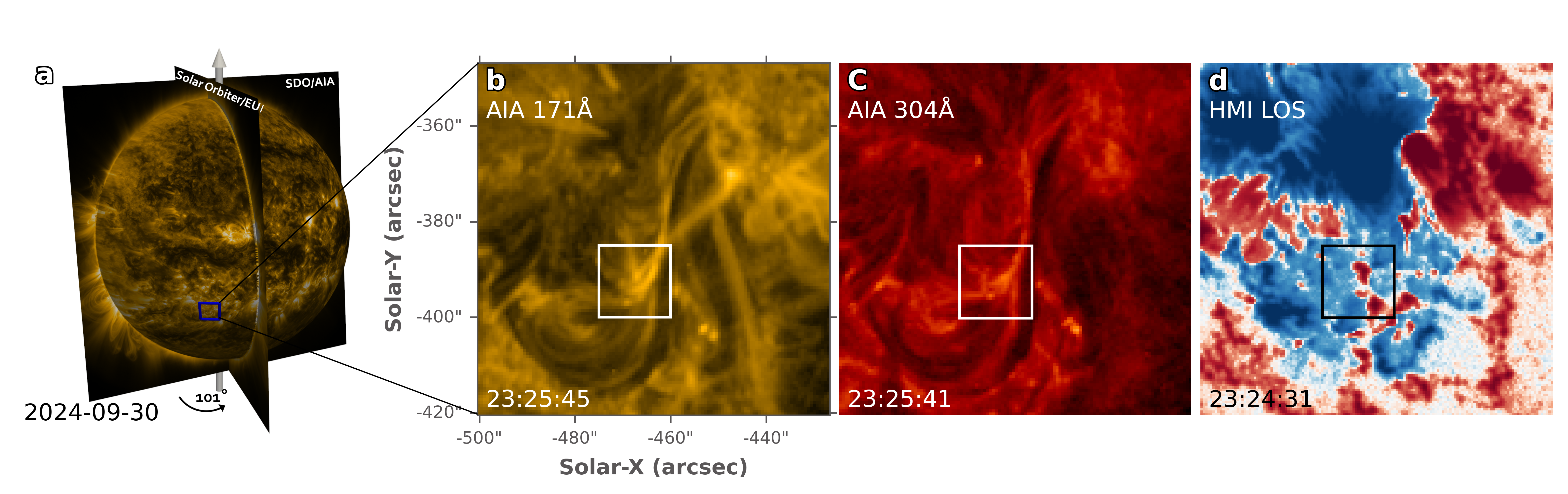}
    \caption{(a) Stereoscopic observation overview with AIA 171 \AA\ and FSI 174 \AA, where the angle between Solar Orbiter and SDO was 101\degr.
    (b) and (c) The filament shown in AIA 171 \AA\ and 304 \AA, displaying a slightly twisted structure with a notable brightening in the central part (indicated by a white box). (d) HMI LOS magnetogram showing the upper section of the filament rooted near a sunspot in positive polarity region. The black box indicates the region corresponding to the brightening in panels (b) and (c), with an evident emerging positive polarity. For the visualisation of the HMI magnetogram, we employed a symmetric logarithmic normalisation (SymLogNorm with linthresh=50, vmin=-750 G, vmax=750 G). Red and blue colours represent positive and negative polarities, respectively.}    \label{figa1}
\end{figure*}

On September 30, 2024, the angle between Solar Orbiter and SDO relative to the Sun was approximately 101\degr, providing a unique opportunity for SDO to observe the limb-erupting filament from a frontal perspective. Fig. \ref{figa1}(a) shows stereoscopic observations from FSI and AIA, while Fig. \ref{figa1}(b) and (c) reveal the entire filament's S-shaped structure with slight twist. A notable brightening appears in the middle section of the filament in both AIA 171 \AA\ and 304 \AA\ images, corresponding to the first enhancement in the STIX light curve. The HMI LOS magnetogram clearly shows positive magnetic flux emergence at the location of the brightening. This paper focuses on jet phenomena observed by $\mathrm{HRI_{EUV}}$. Future works will combine STIX imaging spectroscopy and SDO multi-wavelength data to investigate energy release and particle acceleration mechanisms during this filament eruption.

Due to Solar Orbiter's closer proximity to the Sun (0.29 AU), light from solar eruptions reaches Solar Orbiter approximately 353 seconds before reaching SDO. Thus, the AIA observation in Fig. \ref{figa1} corresponds to the moment when $\mathrm{HRI_{EUV}}$ observation starts, as shown in the accompanying movie.

\begin{figure*}
    \centering
    \includegraphics[scale=0.73]{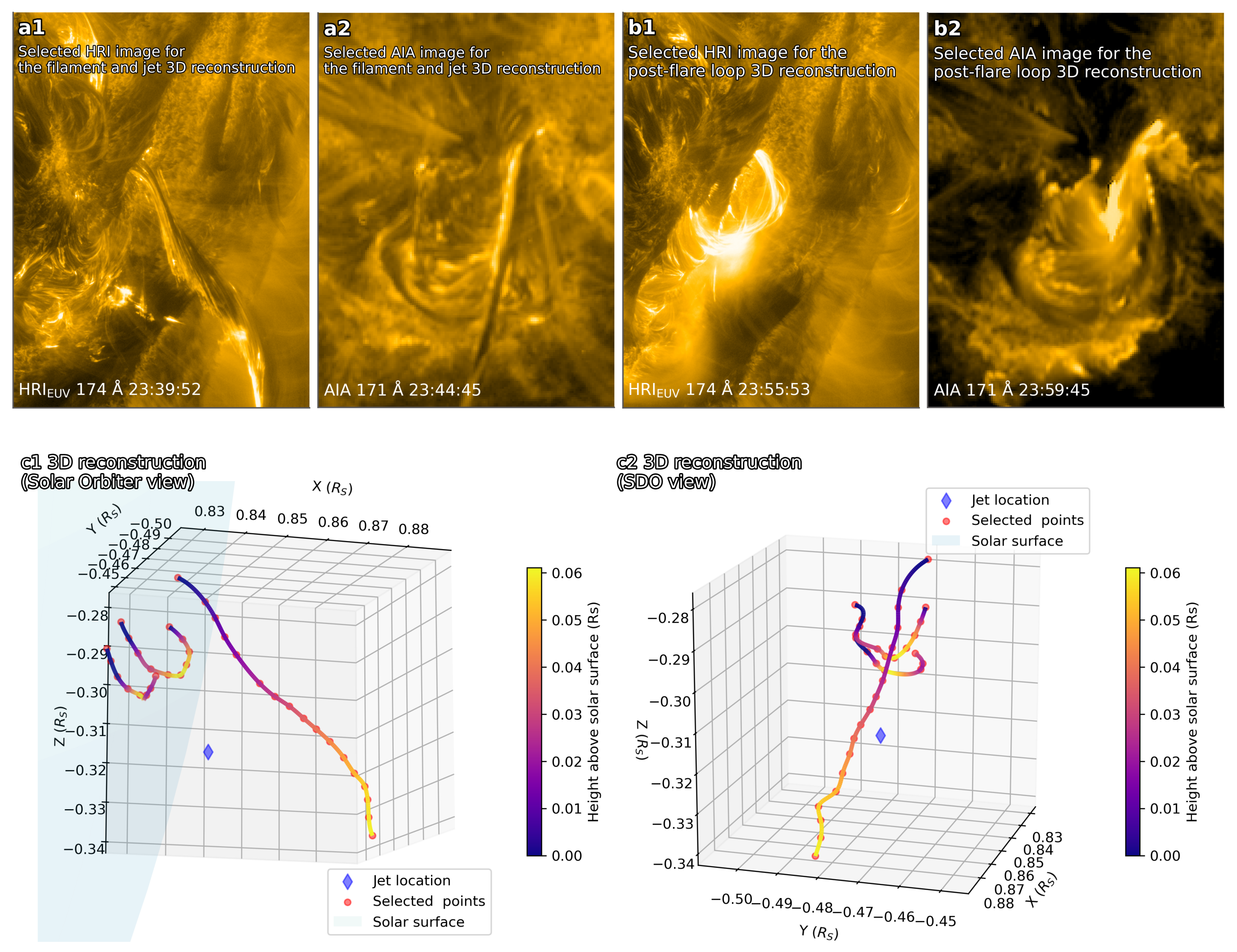}
    \caption{Paired HRI and AIA 171 images used for reconstructing the erupting filament and jet location, shown in (a1) and (a2). Accounting for light travel time differences, $\mathrm{HRI_{EUV}}$  images have been rotated to a certain angle to facilitate
analysis and readability.  Paired $\mathrm{HRI_{EUV}}$ and AIA 171 images used for reconstructing the post-flare loops, shown in (b1) and (b2).  3D reconstruction of the filament, the jet loction, and post-flare-loops from Solar Orbiter and SDO views, respectively, shown in (c1)
and (c2). The points selected for reconstruction are specifically highlighted, with a colour bar indicating the height information of the filament and post-flare-loops.}
    \label{figa2}
\end{figure*}

\section{3D reconstruction of the spatial relationship between filament-jet-post-flare loops }
\label{A2}
The group 2 jets we report differ from those in groups 1 and 3, as they originate beneath the rising filament and exhibit spatial co-location with post-flare loops on top in HRI images. We propose that group 2 jets likely originate from magnetic reconnection at the bottom of the overlying field lines, which are stretched by the erupting filament. However, observations from HRI's single viewing angle may be subject to significant projection effects. Therefore, we reconstruct the erupting filament, the jet location, and the post-flare-loops using stereoscopic observations from paired HRI and AIA 171 Å images. Panels (c1) and (c2) of Fig. \ref{figa2} show the reconstruction results as viewed from the Solar Orbiter and SDO angles, respectively. For details on the 3D reconstruction, we refer to  a previous work in the field \citep{2023SoPh..298...36N}.

Through analysis of the 3D reconstruction results, we obtained the 3D position of group 2 jets (with jet 4 as a representative) relative to the erupting filament and post-flare loops. As depicted in the cartoon diagram in Fig. \ref{fig3} (d), the jet and the post-flare-loops top exhibit co-spatial characteristics and are clearly located below the erupting filament. This spatial configuration leads us to conclude that the group 2 jets were produced by magnetic reconnection caused by the erupting filament stretching the overlying magnetic field. 

However, our interpretation is limited by having only single-wavelength $\mathrm{HRI_{EUV}}$ data. Additionally, the side-view perspective restricted our ability to completely capture additional reconnection evidence below the filament. Therefore, while our explanation is consistent with the observed features, alternative mechanisms cannot be entirely ruled out.

\section{Data availability and jet analysis}
\label{A3}

Solar Orbiter data is publicly available through the Solar Orbiter Archive\footnote{https://soar.esac.esa.int/soar/}. This research used the SunPy \citep{sunpy_community2020,Mumford2020} and \href{https://github.com/mdolab/niceplots}{NicePlots} open software packages to present the observation results. Specifically, we also used \href{https://github.com/sunpy/sunkit-pyvista}{sunkit-pyvista} to visualise the stereoscopic view of the Solar Orbiter and SDO (Fig. \ref{A1}(a)).

This EUI and STIX joint observation were part of a Solar Orbiter major flare campaign that included several observing windows in 2024 (See \cite{2020A&A...642A...3Z,2025arXiv250507472R} for an overview of the campaign). The $\mathrm{HRI}_{\mathrm{EUV}}$ operates with long-exposure images (cadence of 2 seconds). We used 1083 frames of long-exposure images (23:20-23:55 UT) from the $\mathrm{HRI}_{\mathrm{EUV}}$ level-2 data \citep{euidatarelease6}. The cross-correlation technique (see \cite{2022A&A...667A.166C} for details) was applied to reduce the jitter.

The jets we report are characterised by short lifetimes and complex morphologies, making time-space plots potentially inadequate for capturing details. Therefore, we analyzed each jet frame by frame. We placed slices along jets between two frames where jet morphology was relatively complete. We defined jet edges at points where intensity dropped to three-quarters of the maximum-to-minimum range, thus measuring jet displacement. Jet velocity was calculated using the time difference between frames. Simultaneously, slices placed at jet centers, allowed single-peak Gaussian fitting of intensity distributions, with full width at half maximum representing jet width. Notably, due to the highly dynamic nature of the jets and instances where jets decay and re-enhance, the jet velocity properties we obtained may contain significant uncertainty depending on the selection of frames. We calculated each jet's lifetime by counting frames in which it maintained its basic morphology. Additionally, because some jets exhibited fragmentation, we did not analyze jet length properties.

\begin{figure}
    \centering
    \includegraphics[scale=0.58]{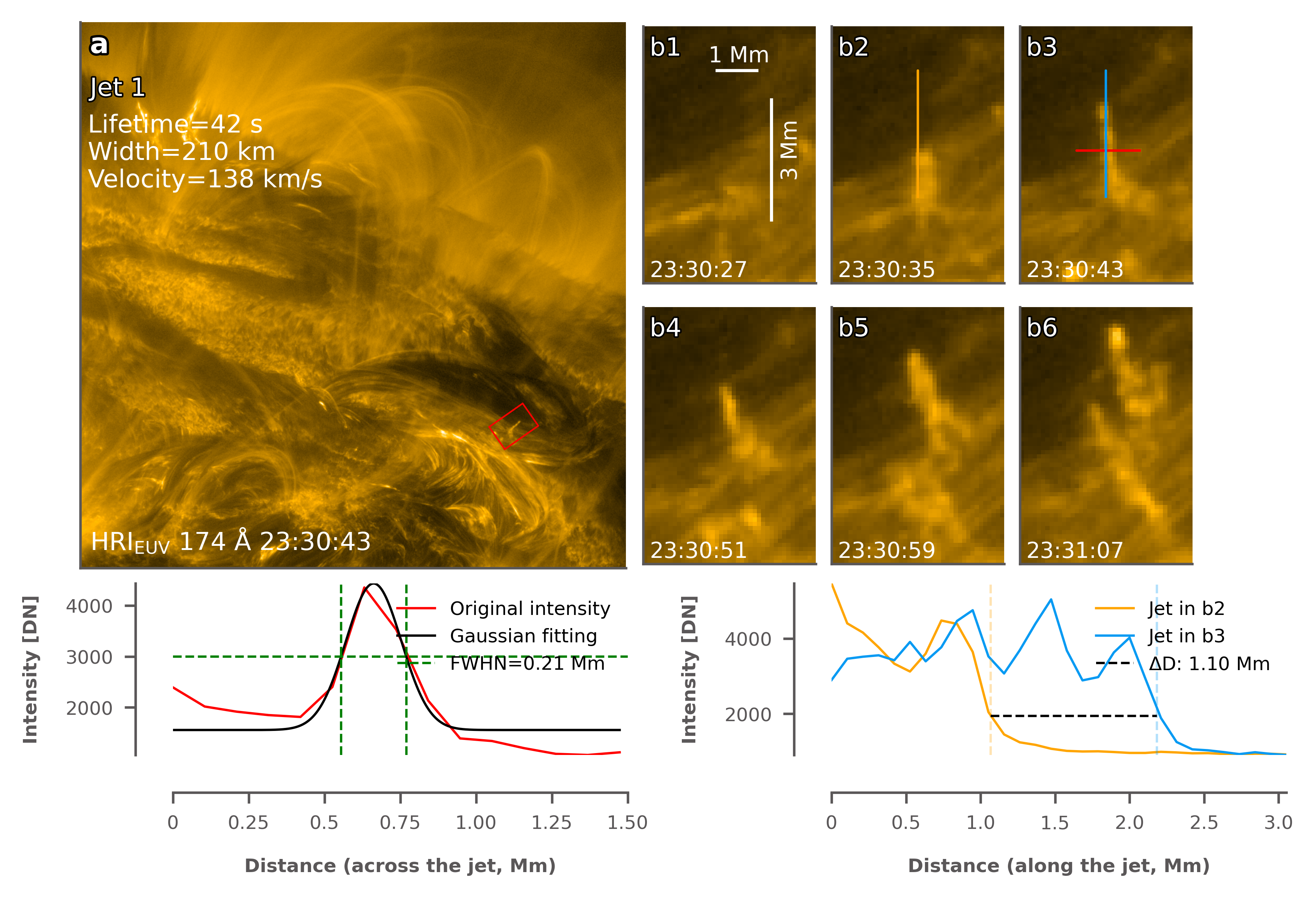}
    \caption{(a): Overview of the erupting filament, marking the location of jet 1, which can be seen originating from the filament spine. (b): Six selected frames with scale bar indicated. By calculating the displacement between b3 and b2 and considering the observation time difference, the velocity of jet 1 is determined to be 138 km/s.}
    \label{J1}
\end{figure}

\begin{figure}
    \centering
    \includegraphics[scale=0.58]{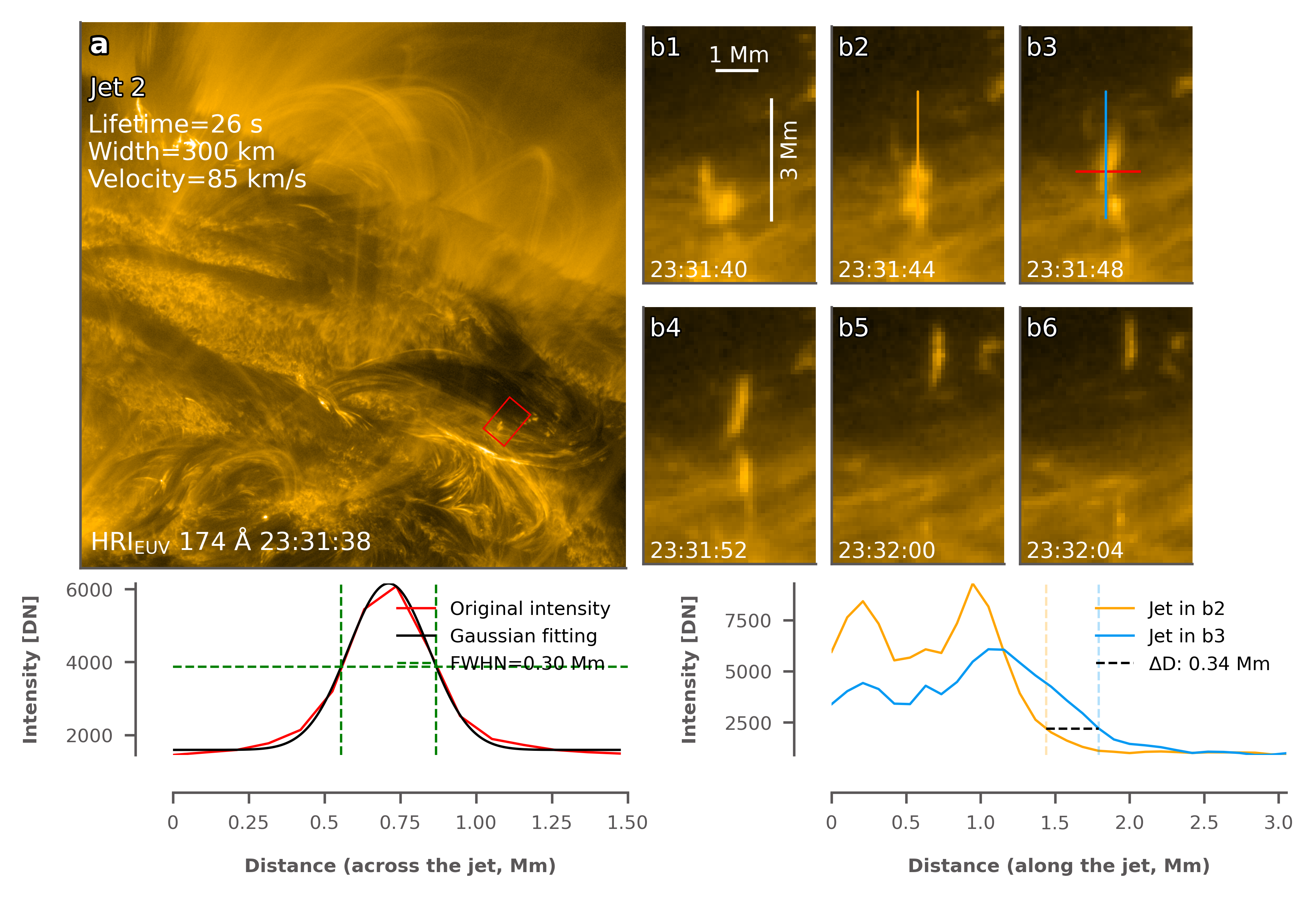}
    \caption{As in Fig. \ref{J1}, but showing jet 2 which originates from the filament spine. Its velocity is 85 km/s}
    \label{J2}
\end{figure}

\begin{figure}
    \centering
    \includegraphics[scale=0.58]{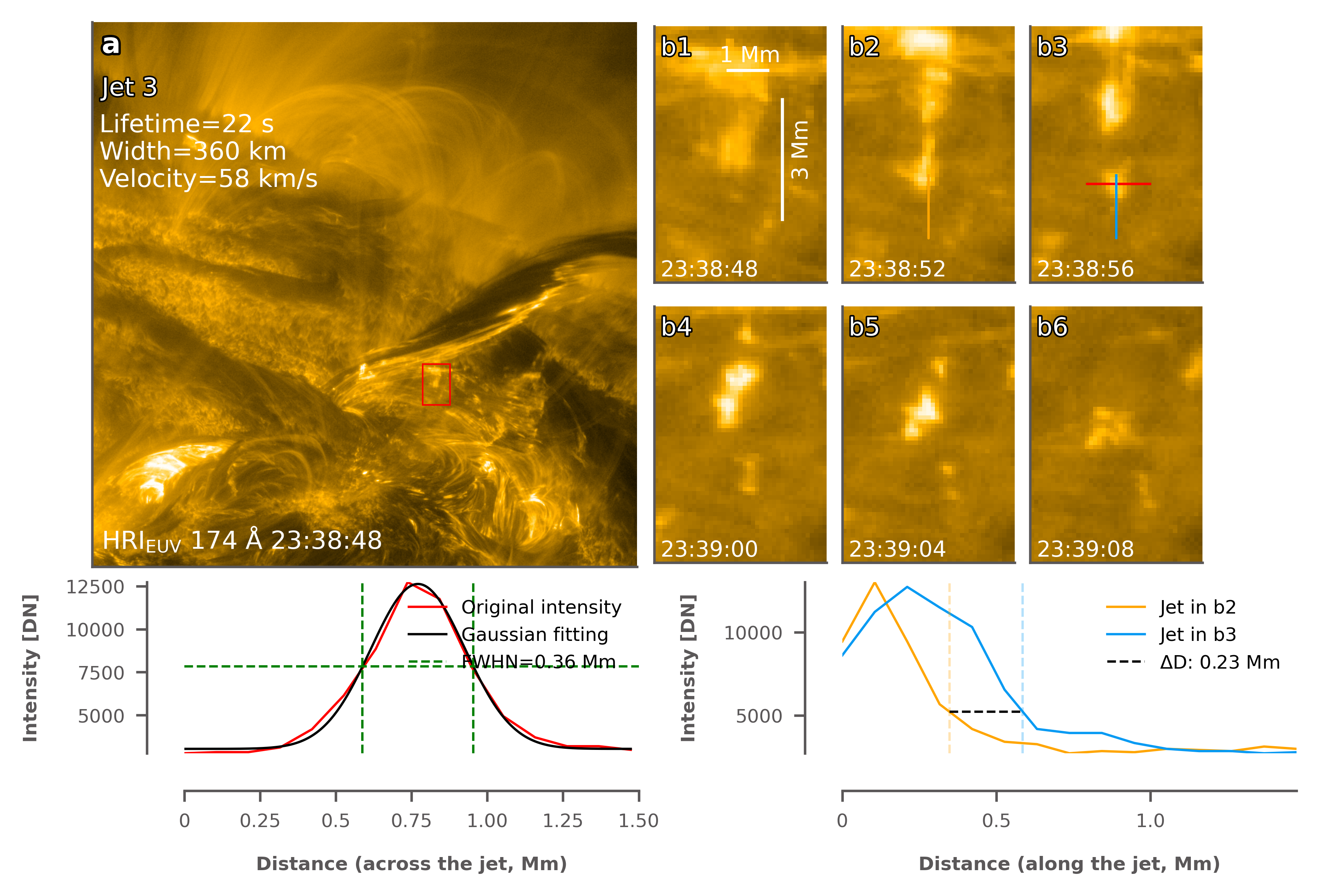}
    \caption{As in Fig. \ref{J1}, but showing jet 3 which originates below the filament spine. Its velocity is 85 km/s}
    \label{J3}
\end{figure}

\begin{figure}
    \centering
    \includegraphics[scale=0.58]{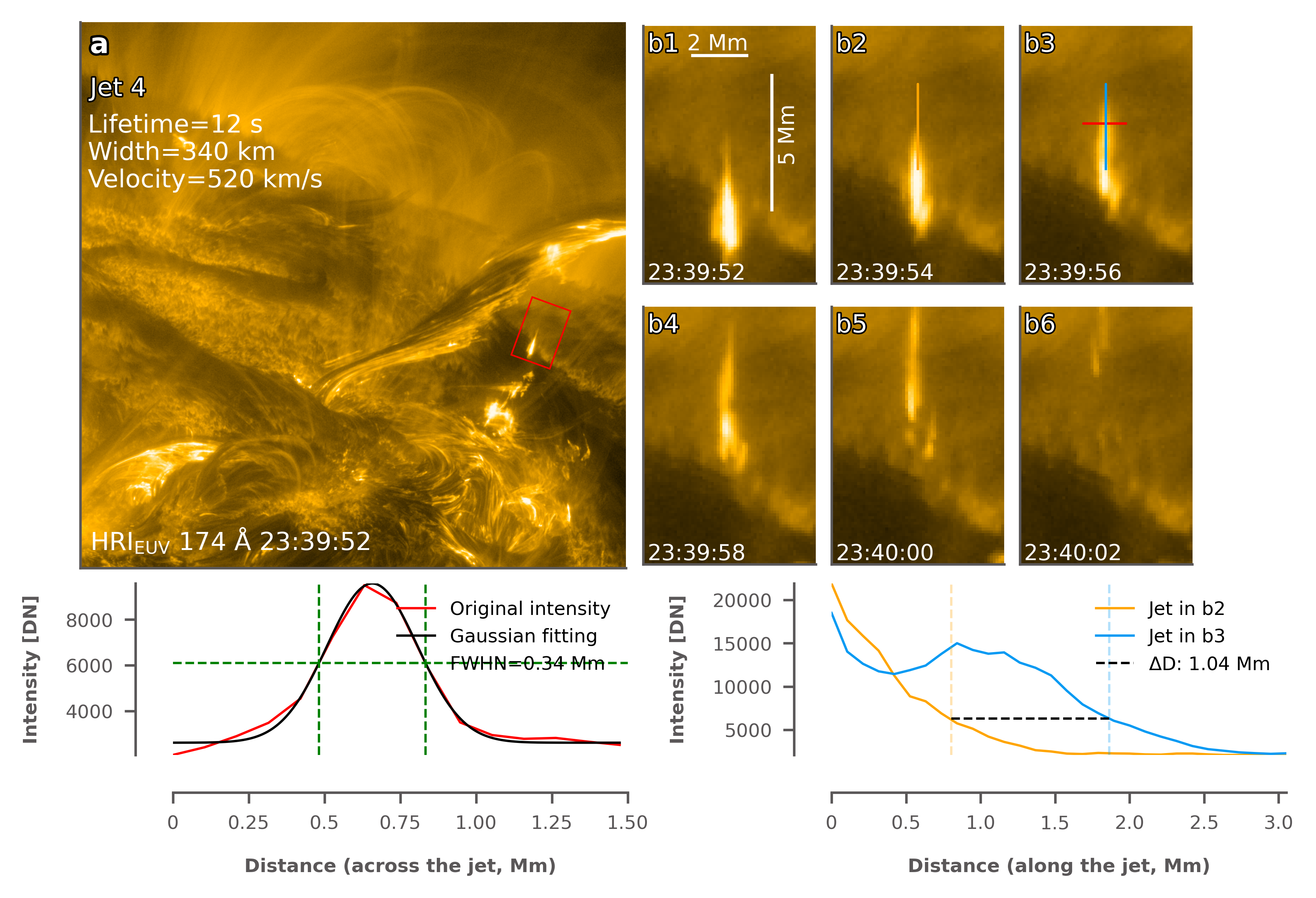}
    \caption{As in Fig. \ref{J1}, but showing jet 4 which originates below the filament spine. Its velocity is 520 km/s.}
    \label{J4}
\end{figure}

\begin{figure}
    \centering
    \includegraphics[scale=0.58]{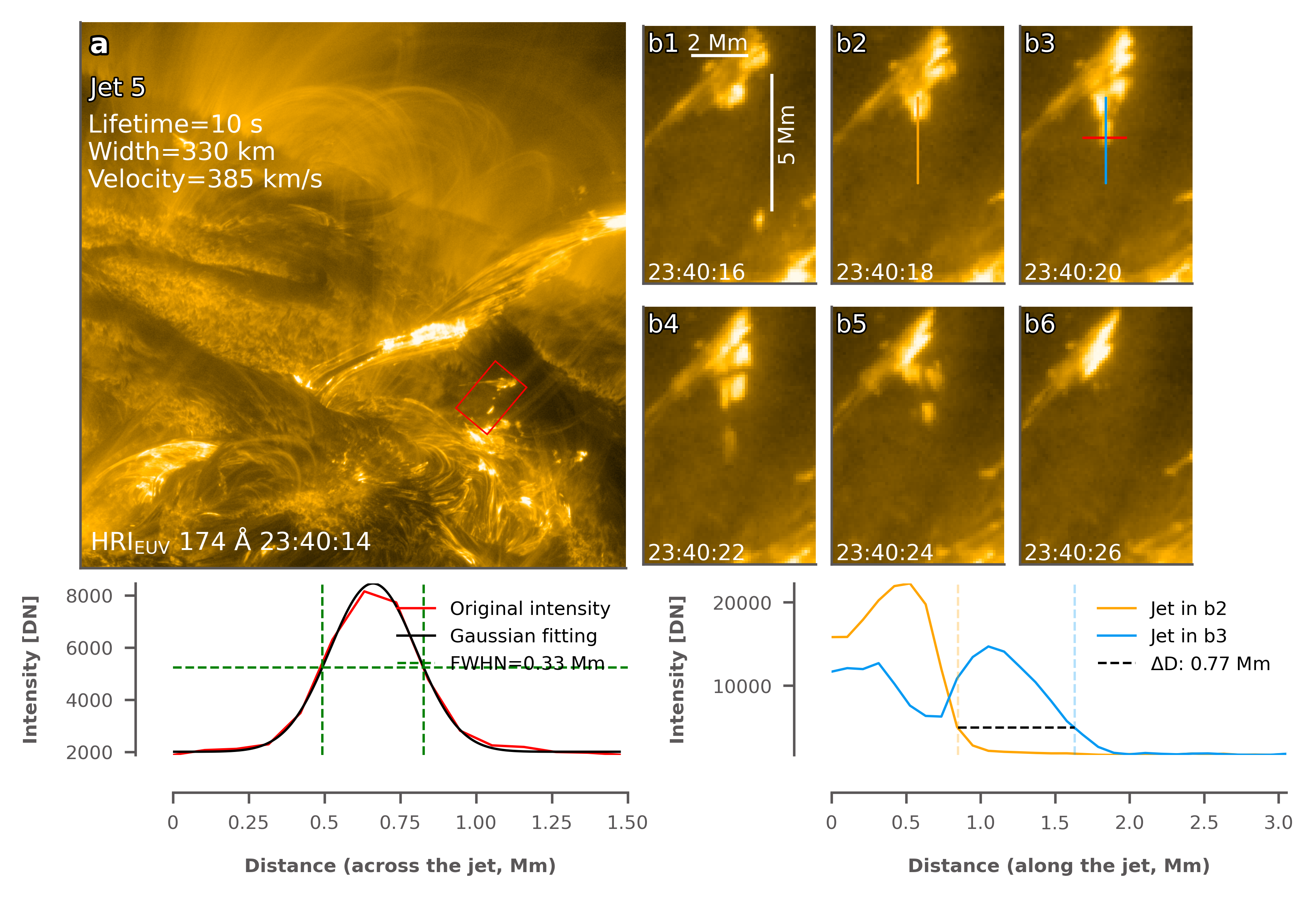}
    \caption{As in Fig. \ref{J1}, but showing jet 5 which originates below the filament spine. Its velocity is 385 km/s}
    \label{J5}
\end{figure}

\begin{figure}
    \centering
    \includegraphics[scale=0.58]{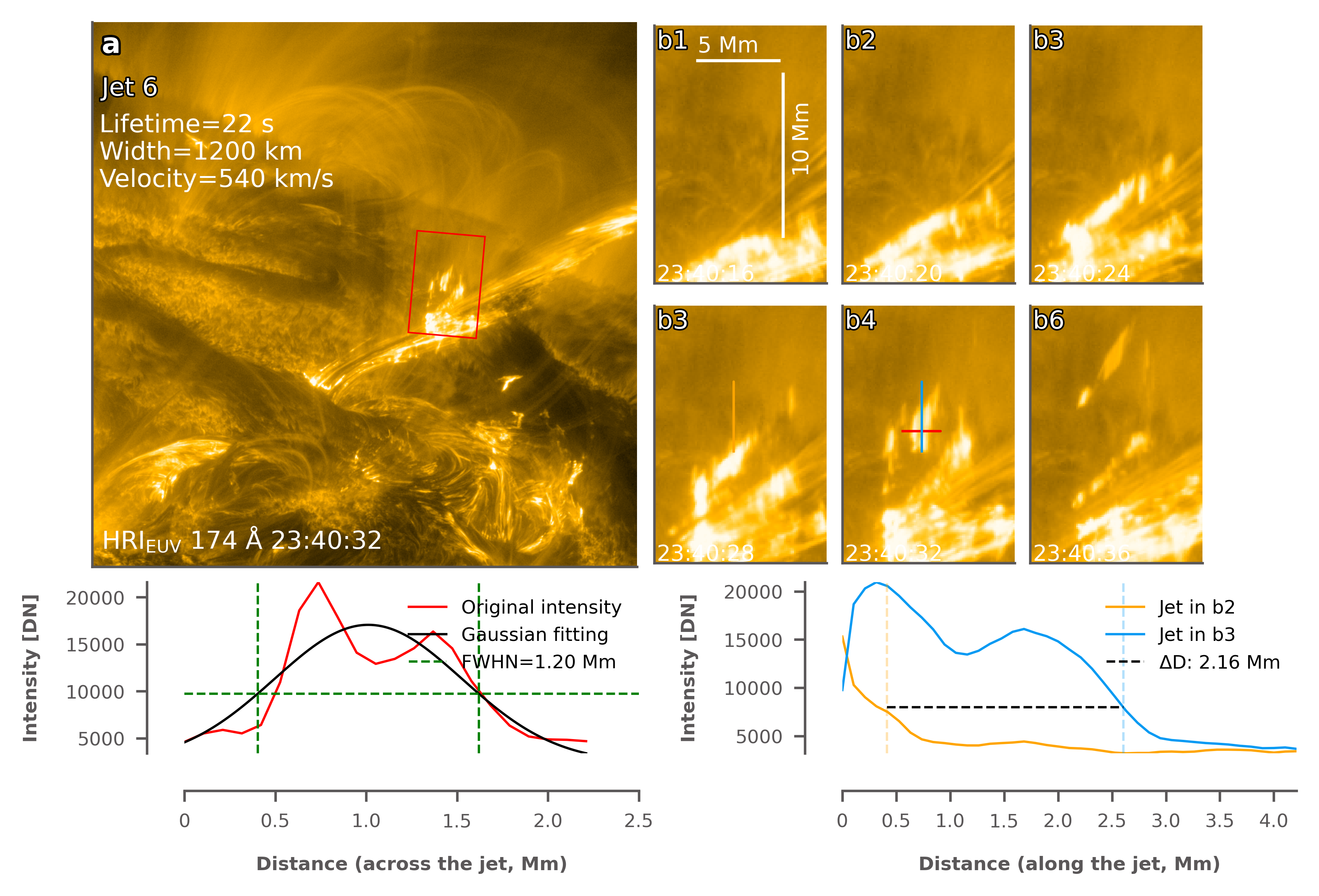}
    \caption{As in Fig. \ref{J1}, but showing jet 6 which originates from the filament spine. Its velocity is 540 km/s}
    \label{J6}
\end{figure}

\begin{figure}
    \centering
    \includegraphics[scale=0.58]{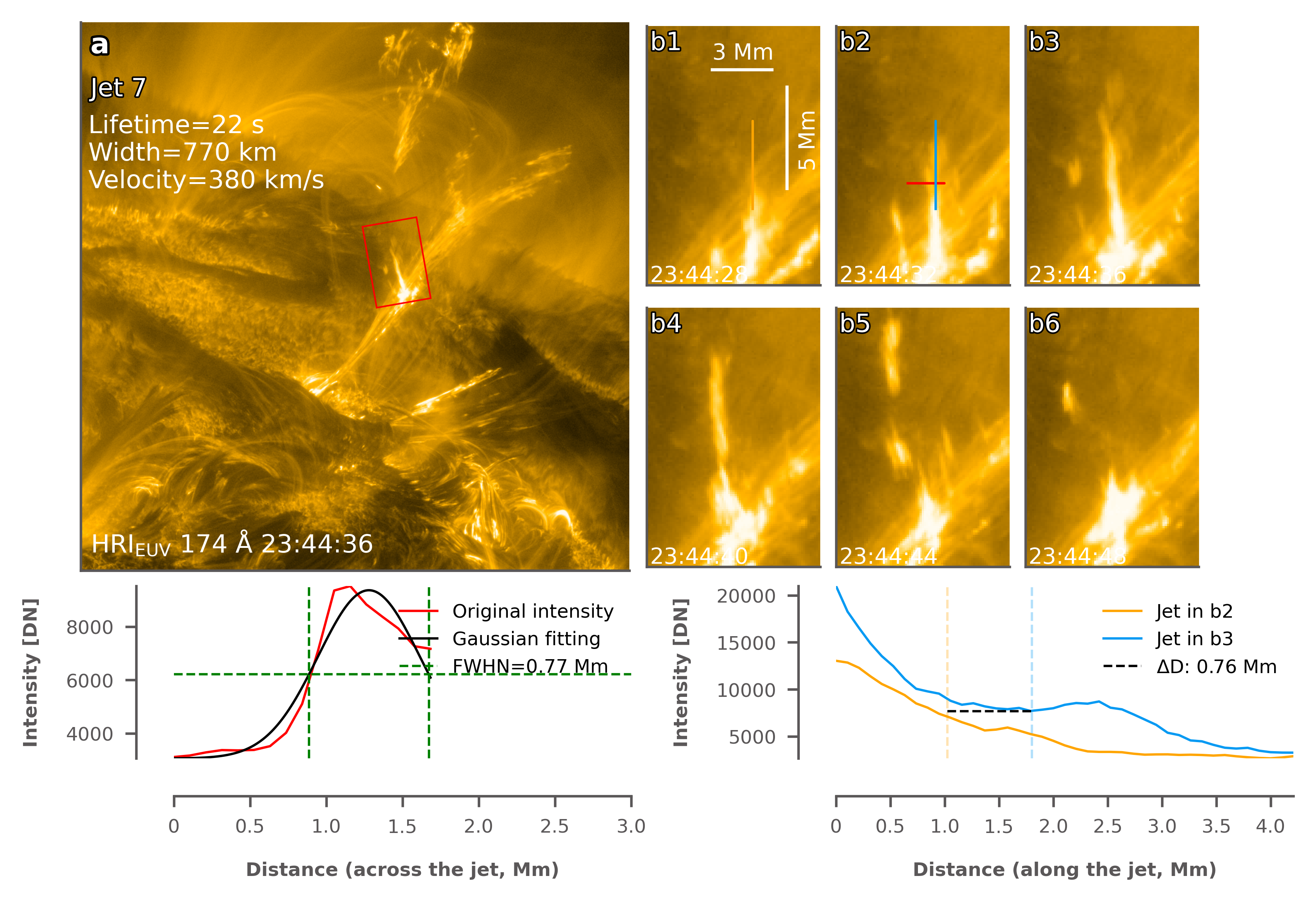}
    \caption{As in Fig. \ref{J1}, but showing jet 7 which originates from the filament spine. Its velocity is 380 km/s}
    \label{J7}
\end{figure}

\begin{figure}
    \centering
    \includegraphics[scale=0.58]{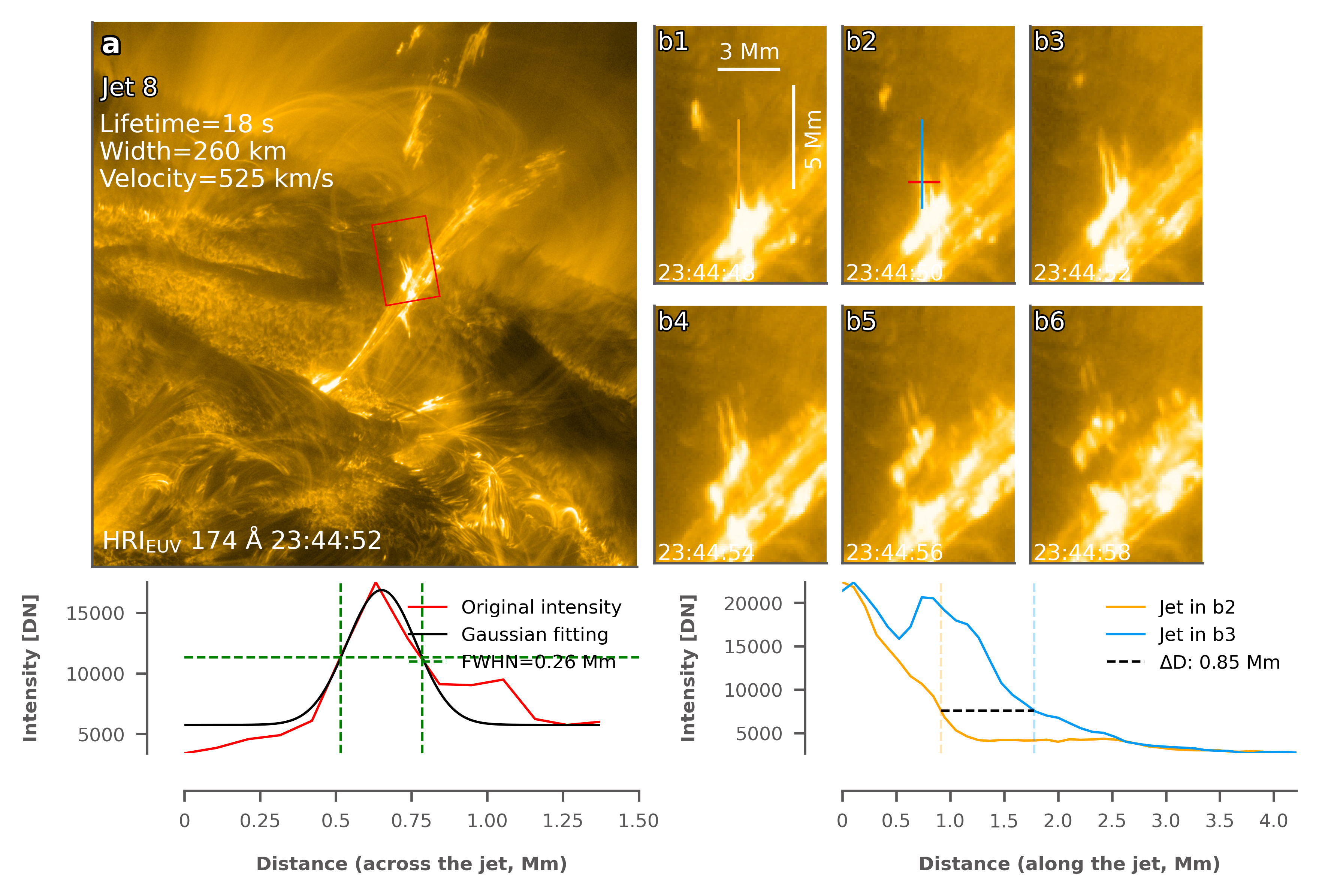}
    \caption{As in Fig. \ref{J1}, but showing jet 8 which originates from the filament spine. Its velocity is 525 km/s}
    \label{J8}
\end{figure}

\begin{figure}
    \centering
    \includegraphics[scale=0.58]{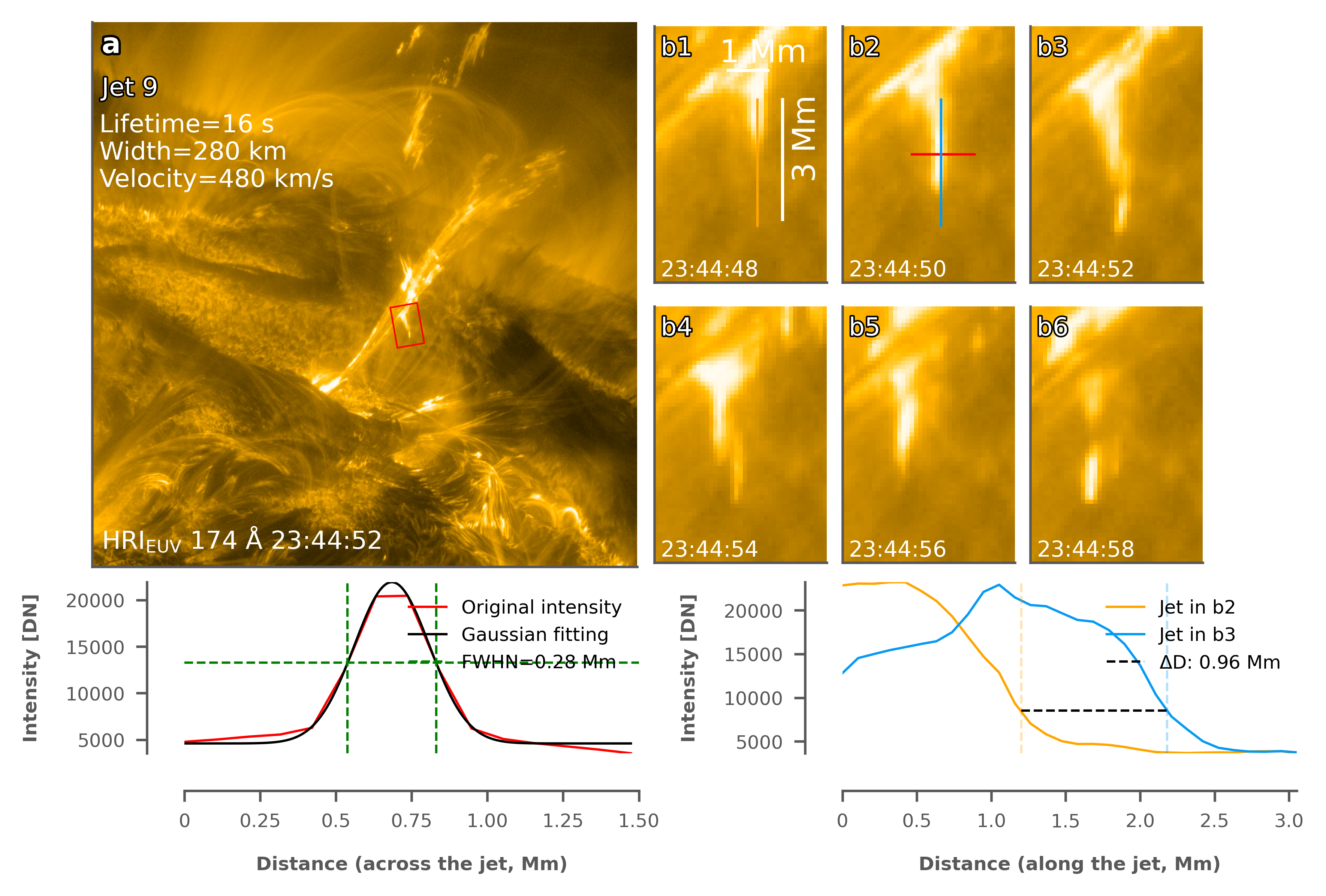}
    \caption{As in Fig. \ref{J1}, but showing jet 9 which originates from the filament spine. Its velocity is 480 km/s}
    \label{J9}
\end{figure}

The analysis results of the nine jets are presented in Figs.~\ref{J1}-\ref{J9}. All jet images were rotated to optimise orientation for analysis and improve readability of the features.


\end{appendix}

\end{document}